\documentclass[fleqn,usenatbib]{mnras}

\usepackage{amsmath}	
\usepackage{mathptmx}
\usepackage{txfonts}

\usepackage[T1]{fontenc}
\usepackage{ae,aecompl}

\usepackage{graphicx}	

\usepackage{amssymb}	

\graphicspath{ {Images/} }

\title[Turbulent driving mode of ionizing radiation]{On the turbulence driving mode of expanding HII regions}

\author[S.~H.~Menon, C.~Federrath and R.~Kuiper]{
Shyam H.~Menon,$^{1}$\thanks{E-mail: Shyam.Menon@anu.edu.au }
Christoph Federrath,$^{1}$
Rolf Kuiper$^{2}$
\\
$^{1}$Research School of Astronomy and Astrophysics, Australian National University, Canberra, ACT~2611, Australia\\
$^{2}$Institute of Astronomy and Astrophysics, University of T{\"u}bingen, Auf der Morgenstelle 10, D-72076 T{\"u}bingen, Germany
}

\date{Accepted XXX. Received YYY; in original form ZZZ}

\pubyear{2020}

\begin{document}
\label{firstpage}
\pagerange{\pageref{firstpage}--\pageref{lastpage}}
\maketitle

\begin{abstract}
We investigate the turbulence driving mode of ionizing radiation from massive stars on the surrounding interstellar medium (ISM). We run hydrodynamical simulations of a turbulent cloud impinged by a plane-parallel ionization front. We find that the ionizing radiation forms pillars of neutral gas reminiscent of those seen in observations. We quantify the driving mode of the turbulence in the neutral gas by calculating the driving parameter $b$, which is characterised by the relation $\sigma_s^2 = \ln({1+b^2\mathcal{M}^2})$ between the variance of the logarithmic density contrast $\sigma_s^2$ (where $s = \ln({\rho/\rho_0})$ with the gas density $\rho$ and its average $\rho_0$), and the turbulent Mach number $\mathcal{M}$. Previous works have shown that $b\sim1/3$ indicates solenoidal (divergence-free) driving and $b\sim1$ indicates compressive (curl-free) driving, with $b\sim1$ producing up to ten times higher star formation rates than $b\sim1/3$. The time variation of $b$ in our study allows us to infer that ionizing radiation is inherently a compressive turbulence driving source, with a time-averaged $b\sim 0.76 \pm 0.08$. We also investigate the value of $b$ of the pillars, where star formation is expected to occur, and find that the pillars are characterised by a natural mixture of both solenoidal and compressive turbulent modes ($b\sim0.4$) when they form, and later evolve into a more compressive turbulent state with \mbox{$b\sim0.5$--$0.6$}. A virial parameter analysis of the pillar regions supports this conclusion. This indicates that ionizing radiation from massive stars may be able to trigger star formation by producing predominately compressive turbulent gas in the pillars.

\end{abstract}

\begin{keywords}
MHD -- turbulence -- ISM -- star-formation -- HII-regions
\end{keywords}



\section{Introduction}
\label{sec:introduction}

Stars predominantly form in dense, gravitationally-bound gas inside giant molecular clouds (GMCs). Massive stars or clusters (OB associations) in these clouds produce UV radiation that dissociates and photoionizes the surrounding gas and results in the expansion of an ionization front (HII region). Photoionization increases the thermal pressure within GMCs, has the ability to mechanically unbind and destroy them \citep{Williams_1997,Matzner_2002}, and drives turbulence that may control the formation of the next generation of stars \citep{Elmegreen_2004,Mac_2004,McKee_2007,federrath_klessen_2012,Padoan_2014,Krumholz_2019}. 

The expanding HII regions are also known to sculpt the surrounding neutral gas into structures reminiscent of the iconic 'Pillars of Creation' imaged by the Hubble Space Telescope \citep{Hester_1996}. Since then, there have been a wealth of observations using multiwavelength surveys that image these pillars and related structures such as globules, energetic evaporating globules (EEG's) and proplyds, and deduce dynamical quantities in and around them \citep{Preibisch_2012,Klaassen_2014,Hartigan_2015,Schneider_2016,Klaassen_2018,Klaassen_2019}. Models proposed to explain their formation lie broadly in two categories: the classic collect-and-collapse model by \cite{Elmegreen_1977} where the HII region sweeps up and accumulates cold gas creating density enhancements and eventually pillars in their shadows; or the more recent radiation-driven implosion (RDI) model where clouds with pre-existing density enhancements are sculpted to form pillars by impinging ionizing radiation. In these studies the density enhancements are modelled as Bonnor-Ebert spheres or are seeded by turbulence, and they naturally produce the observed morphologies and dynamics of pillars \citep{Mellema_2006,Gritschneder_2010,Mackey_2010,Walch_2012,Tremblin_2013}. 

Direct signatures of star formation are observed at the tip of these pillars through jets and outflows with suggestions that this star formation could be triggered by the ionizing radiation \citep{Sugitani_2002,Billot_2010,Smith_2010,Chauhan_2011,Reiter_2013,Cortes-Rangel_2019}. However, the RDI picture of forming pillars raises the question as to whether star formation is really 'triggered', as stars could have formed by the direct gravitational collapse of the pre-existing density enhancements anyway, and the ionizing radiation need not have enhanced local star formation in any way. In addition, numerical simulations have shown that photoionization may actually have a negative global net effect on star formation in GMCs, helping to explain the observed low star formation rates \citep{Vazquez_Semadeni_2010,Dale_2012,Colin_2013,Gavagnin_2017,Geen_2017,Kim_2018}. However, photoionization feedback may also simultaneously trigger star formation \emph{locally}, by increasing the number of stars formed (but not necessarily the total mass of stars) \citep{Dale_2013}. Moreover, \cite{Dale_2015_Trigger} argues that current observational markers used to infer triggering may not be robust enough to distinguish whether an observed star was triggered or has spontaneously formed. Thus, there is no consensus on whether triggered star formation is effective on GMC scales, and hence resolving and understanding the local dynamics of the gas, in particular the turbulent flow is crucial for determining the role of photoionization for star formation. 

Turbulence regulates star formation in molecular clouds and can play a dual role in the process: providing a form of support against self-gravitational collapse due to random velocity fluctuations, and on the other hand forming shocks resulting in overdensities that eventually undergo gravitational collapse \citep{Mac_2004}. However, the observed supersonic turbulence \citep{Elmegreen_2004} must be driven by some external mechanism, as it would otherwise decay within one crossing time \citep{Mac_Low_1998,Mac_2004,Stone_1998}. Various agents for this driving have been proposed such as gravity, accretion, galactic rotation, shearing flows, stellar feedback, etc.~\citep[see][for a review of potential drivers]{Federrath_2017,Federrath_2018}, with the difference lying in the type of turbulent modes that they drive: compressive (curl-free) or solenoidal (divergence-free) modes. This difference is quantified through the driving parameter $b$, which typically varies between 1/3 and 1, where these two extreme cases refer to purely solenoidal and purely compressive driving, respectively \citep{Federrath_2010}. The value of $b$ is important, as the flow dynamics, density structure and the subsequent star formation rate depend on it \citep{Federrath_2008,Federrath_2010,Price_2011,Konstandin_2012,Padoan_2014,Federrath_2015,Nolan_2015}; with compressive driving resulting in broader density probability distribution functions (PDFs) and star formation rates approximately an order of magnitude larger than for solenoidal driving \citep{federrath_klessen_2012,Federrath_2016,Federrath_2018}. The values of $b$ have been studied systematically for different driving sources of turbulence in numerical simulations \citep{Pan_2016,Kortgen_2017,Jin_2017}, and observations also find a significant variation in $b$ across different clouds in the Milky Way \citep{Padoan_1997,Brunt_2010,Ginsburg_2013,Kainulainen_2013,Federrath_2016,Kainulainen_2017}.

Expanding HII regions have been proposed to be one of the primary sources of driving and maintaining supersonic turbulence in GMCs \citep{Matzner_2002,Krumholz_2006,Goldbaum_2011}. For instance, \cite{Gritschneder2009} simulate the ionizing irradiation of a local patch of a cloud ($\sim4\,\mathrm{pc}$) and compare it to control runs without irradiation where the the initial turbulence is allowed to decay. They measure compressive, solenoidal and total power spectra and find that turbulence is driven significantly in the cold neutral gas, particularly in compressive modes, with more efficient driving on smaller scales, leading to a flatter power spectrum. However, there has been no systematic study of the driving parameter $b$ based on density and velocity fluctuations for turbulence driven by expanding HII regions. Studying this would allow for more direct inferences on whether the local gas dynamics inside and around pillars shaped by the ionizing radiation support a picture of triggered star formation or not.

In this study we calculate the driving parameter $b$ of ionizing radiation-driven turbulence on the neutral gas in the vicinity of expanding HII regions. We perform numerical simulations analogous to \cite{Gritschneder2009}, wherein we model the incoming radiation front as plane-parallel, and then calculate $b$ from the previously established relation \citep{Padoan_1997,Federrath_2008,Federrath_2010,Price_2011,Padoan_2011,Konstandin_2012,Molina_2012,Hopkins_2013,Federrath_2015,Nolan_2015,Squire_2017,Mandal_2019},
\begin{equation} \label{eq:sigmas}
\sigma_s^2 = \ln\left(1+b^2\mathcal{M}^2\right),
\end{equation}
where $\sigma_s$ is the standard deviation of the logarithmic density contrast $s=\ln(\rho/\rho_0)$ and the turbulent, sonic rms Mach number ($\mathcal{M}$). We further calculate the value of $b$ for sub-parsec-scale regions at the tip of the formed pillar-like structures, where star formation is expected to occur.

In Section~\ref{sec:methods} we define the simulation setup, initial condition, and the numerical methods we use in the simulations. In Section~\ref{sec:results} we present our results and Section~\ref{sec:summary} provides a summary and conclusions.

\section{Numerical methods}
\label{sec:methods} 

\subsection{Physics and numerical methods}

To model the evolution of the cloud and ionizing radiation, we follow the equations of compressible three-dimensional hydrodynamics in combination with an incoming plane-parallel photoionizing radiation flux in the $x$-direction of the Cartesian computational domain. For the hydrodynamics, we utilize the open source code \emph{Pluto} \citep{PLUTO_2007,PLUTO_2012} in version 4.1. For the photoionization feedback, we make use of the \emph{Sedna} module, a ray-tracing radiation transport solver, introduced in \citet{Kuiper_2018}. This solver combination was also used for a recent code benchmark \citep{Starbench}. To generate the initial turbulent ISM setup, we implement the turbulence generator described in \citet{Federrath_2008} and \citet{Federrath_2010}.

\subsection{Initial Conditions}
Our simulation setup is largely similar to that of the fiducial simulation in \cite{Gritschneder2009}. The initial setup is a uniform Cartesian grid of length $L = 4 \, \mathrm{pc}$ containing $1000 \, \mathrm{M}_\odot$ of neutral gas at a temperature of $T_\mathrm{neutral} = 10 \, \mathrm{K}$ corresponding to a sound speed of $c_\mathrm{s} = 0.28 \,\mathrm{km}\,\mathrm{s}^{-1}$. To initialise a neutral turbulent state, we introduce supersonic velocity fluctuations with an rms Mach number of 10, i.e., $\mathcal{M} = \sigma_v / c_\mathrm{s} = 10$, the ratio of turbulent velocity dispersion and sound speed, with a velocity power spectrum following $E(k) \propto k^{-2}$  \citep{Federrath_2010,Federrath_2013}, consistent with supersonic molecular cloud turbulence \citep{Larson_1981,Solomon_1987,Ossenkopf_2002,Brunt_2002,Roman-Duval_2011}. These velocity fluctuations are introduced in the wavenumber range $2 \leq k/(2\pi/L) \leq 20$, with a natural mixture of solenoidal and compressive modes \citep[$b\sim0.4$; see Fig.~8 in][]{Federrath_2010}. This is enforced by applying a projection in Fourier space, which decomposes the driving field into its solenoidal and compressive components \citep{Federrath_2008, Federrath_2010}. We then allow the turbulence to develop self-consistently by evolving the gas hydrodynamically with an isothermal equation of state for one turbulent crossing time given by $t_\mathrm{crossing}  = L/\sigma_v$, where $\sigma_v$ is the velocity dispersion. The computational volume boundaries are periodic in all spatial directions during this phase. The turbulent state obtained after $t = t_\mathrm{crossing}$ has a Mach number of $\sim 3.5$ due to the decay of the initial turbulence. We show a column density projection of this stage in Figure~\ref{fig:Column_Density} (top-left panel), which serves as the initial condition for the subsequent evolution of the cloud when it is hit by the ionization front entering from the left-hand side of the $x$-axis. We define this stage as time $t = 0$ from here on.

\subsection{Simulation Setup}

After this turbulent state is reached ($t = 0$), we introduce the ionizing radiation of a massive star onto the turbulent gas. As we are interested in a local region of a molecular cloud at the edge of the Str{\"o}mgren sphere \citep{Stromgren}, we approximate the ionization front as plane-parallel, impinging from the negative $x$-direction in the simulation domain. The photon flux per unit time is set to $\mathrm{F_{Ly}} = 5 \times 10^9 \; \mathrm{photons} \,\mathrm{cm}^{-2} \, \mathrm{s}^{-1}$. The ionization cross section is set to $6.3 \times 10^{-18} \, \mathrm{cm}^2$ and the recombination rate into any state but the hydrogen ground state is set to $2.6 \times 10^{-13}\,\mathrm{cm}^3\,\mathrm{s}^{-1}$, making use of the so-called on-the-spot approximation \citep{Baker_1938,Spitzer_1978,Osterbrock_1989}. The ray tracer sets an ionization degree $\eta$ at each cell from this flux, which is used to calculate the local temperature in the cell by linear interpolation, 
\begin{equation}
T = T_\mathrm{ion} \times \eta + T_\mathrm{neutral} \times (1-\eta),
\end{equation}
where $T_\mathrm{neutral} = 10\,\mathrm{K}$ and $T_\mathrm{ion} = 10^4\,\mathrm{K}$ are the neutral and ionized gas temperatures, respectively \citep{Shu}. Both gas components can be approximated as being in thermal equilibrium, since the heating and cooling timescales are much shorter than the dynamical timescale. The sound speeds $c_\mathrm{s} = \sqrt{k_\mathrm{B}T/(\mu m_\mathrm{H})}$ associated with these phases are $0.28 \, \mathrm{km} \, \mathrm{s}^{-1}$ and $12.84 \, \mathrm{km}\,\mathrm{s}^{-1}$, respectively, for atomic hydrogen gas with $\mu = 1.0$.

The fluid boundary conditions after switching on the radiation field are periodic in all directions except the $x$-direction, where semi-permeable walls are imposed, i.e., we allow the gas to leave the computational domain through these boundaries, but do not allow new gas to enter. However, the expansion of the HII region expels the neutral gas in the cloud through the semi-permeable boundary at $x = 4 \, \mathrm{pc}$, compromising the study of turbulence driving as the gas leaves the computational domain too early. To delay this expulsion from the computational domain we account for the bulk motion of the gas in the positive $x$-direction, by subtracting the instantaneous centre of mass velocity from each cell's local velocity at each integration step. This allows us to follow the evolution of the cloud and pillars in the rest frame of the system and for a sufficiently long time that includes the formation, evolution and destruction (of some of) the pillars.

The adiabatic index ($\gamma$) of the ideal gas is set to $1.0001$, which effectively establishes a locally isothermal state with the local temperature set by the ionization degree of the gas. For simplicity, we assume the gas is purely atomic with the mass per atom $m_\mathrm{H} \, = 1.6 \times 10^{-24} \, \mathrm{g}$. Note that we do not include magnetic fields, self-gravity or a chemical network in our study. The grid resolution was set to $200^3$ grid cells, and a single run took approximately 300 hours on 140 compute cores. However, we also provide a resolution study to compare our main results for a grid resolution of $100^3$, $200^3$, and $400^3$ grid cells, in Appendix~\ref{sec:Resolution_Study}.

We note that the ray tracer solves the radiative transfer equation using the on-the-spot (OTS) approximation. This neglects the effect of the diffuse extreme ultraviolet (EUV) radiation emitted through recombinations in the surrounding ionized gas. This is believed to be responsible for shadows seen behind the pillars formed in other studies as well as our simulations and is an artefact of the uni-directional ray tracing performed in such a simulation setup. This leads to parcels of low-density gas lying in the shadows of these pillars to remain neutral ($T \sim 10\, \mathrm{K}$), that would in reality be at much higher temperatures due to the EUV recombination field from the surrounding ionized gas. Numerical studies that include the diffuse radiation field find that shadow regions with low density are at least partly ionized and as a result the pillar structures are more compressed and less coherent \citep{Ercolano_2011,Haworth_2012}. We account for this effect by excluding computational cells for which the temperature is inaccurate given their density, from the analysis, as explained in the results section below.

\section{Results}
\label{sec:results}

\subsection{Gas structure and evolution}
\label{subsec:gas_structure}

\begin{figure*}
\centering
\includegraphics[width=\textwidth]{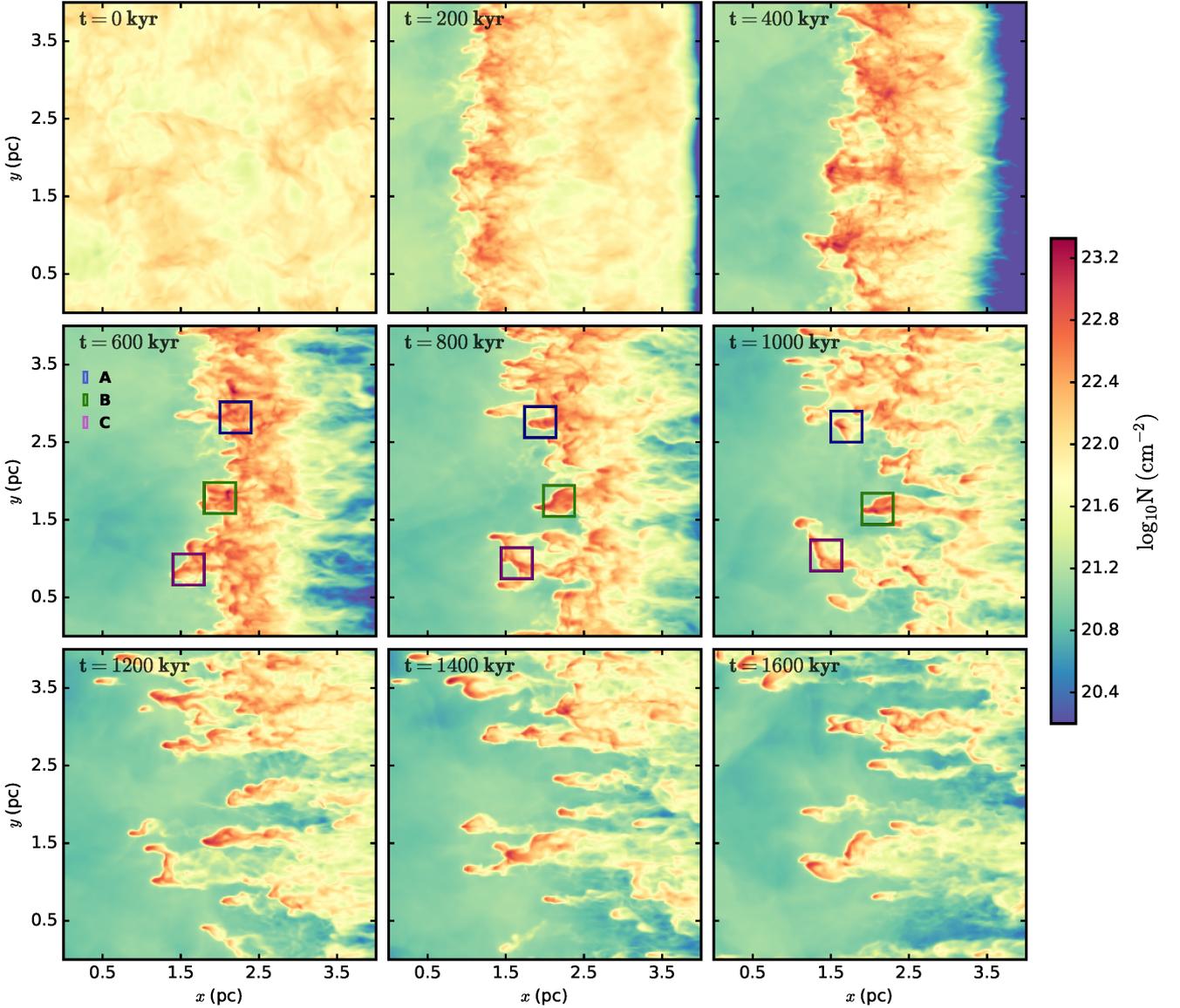}
\caption{Time evolution of the column density of the gas integrated along the $z$-direction. The UV ionizing radiation enters the computational volume from the $x=0$ face. The radiation ionizes parts of the cloud that sculpts pillar-like columns of neutral gas with the densest parts at its tips. The boxes in the 2nd row of panels indicate three sub-parsec scale ($\sim0.4\mathrm{pc}$) regions containing the tips of pillars where we individually analyse the turbulence driving mode.}
\label{fig:Column_Density}
\end{figure*}

The initial turbulent state is shown in Figure~\ref{fig:Column_Density} (top-left panel). The ionization front is impinging onto this turbulent gas from the lower $x$-boundary of the computational domain (i.e., the $x = 0$ face). It instantly ionizes parts of gas in the cloud, with the radiation penetrating further into low-density channels, forming a shock front at the transition from ionized to neutral gas as can be seen at $t=200 \, \mathrm{kyr}$ in the 2nd panel of Figure~\ref{fig:Column_Density}. On further evolution, the radiation continues to penetrate the low-density channels but fails to ionize columns of gas in the $x$-direction that contain overdensities seeded by the initial turbulent field. The ionized channels expand and through their thermal pressure push the neutral gas in their surroundings towards the columns of neutral gas in the shadows of overdensities. This process compresses the gas tangentially to the direction of the radiation and as a result, pillar-like structures form in the cold neutral gas (top-right panel in Figure~\ref{fig:Column_Density}). Further evolution unveils discernible pillar-like structures surrounded by a pool of hot ionized gas reminiscent to those seen in observations (three selected pillar regions are marked in the 2nd row of Figure~\ref{fig:Column_Density}). In addition to this thermal pressure-induced compression, the surrounding ionized gas can push off the neutral material, introducing a back-reaction known as the "rocket effect" that could exert significant forces and excite turbulent modes in the neutral gas \citep{Krumholz_Clusters}. Note again that we are following the evolution in the rest frame of the entire system, which is why some of the pillars appear to move against the radiation direction.

Figure~\ref{fig:Column_Density} shows that the tips of the pillars have the highest density enhancements. It is at these pillar tips that star formation is observed to occur \citep[e.g.,][]{Smith_2010,Reiter_2013,Klaassen_2014,Klaassen_2019}. Thus, we define three different pillar regions for more detailed analysis below: Pillar A, B and C of sizes $0.4\,\mathrm{pc}$ in each direction. The regions are picked by eye from their line-integrated column densities in each direction while ensuring that the high-density tip of the pillar is included, and then adjusted such that the centre of mass of the region lies at the centre of the defined cubical box. These defined regions are shown for the line-integrated column density in the $z$-direction in Figure~\ref{fig:Column_Density} (2nd row and marked with boxes) for three different time snapshots during which we follow their evolution.

At later stages of the evolution ($t>1000\,\mathrm{kyr}$) the ionizing radiation succeeds in photo-evaporating or expelling a large fraction of neutral gas in the cloud. However, following the system in the instantaneous centre of mass frame of reference allows us to see the late time evolution of some pillars and isolated globules that manage to survive. Their survival can be attributed to the strong turbulent ram pressure in their interiors with comparable strengths to the thermal pressure of the ionized gas. Observations have revealed such structures in the vicinity of HII regions and they are usually categorised as globules, evaporating gaseous globules (EGGs), condensations and proplyds based on their morphology and sizes, and are proposed to eventually form from radiation sculpted pillar-like structures \citep[see for instance][]{Schneider_2016}.

\subsection{Driving Parameter $b$}

The driving parameter $b$ is a quantity that is proportional to the ratio of density to velocity fluctuations, $b \propto \sigma_{\rho}/\sigma_{v}$ in a supersonically turbulent cloud \citep{Federrath_2008,Federrath_2010}. A velocity field that contains primarily compressible modes would produce stronger compressions and rarefactions and thus result in a higher spread in the density PDF than a primarily solenoidal velocity field. From the well-studied relation given by Equation~(\ref{eq:sigmas}) between the log-density dispersion $\sigma_s$ and the rms turbulent Mach number $\mathcal{M}$, we can rearrange for the driving parameter $b$,
\begin{equation}
	b = \left\{\left[\exp\left(\sigma_s ^2\right) - 1\right]/\mathcal{M}^2\right\}^{1/2}.
	\label{eq:b_parameter}
\end{equation}
Calculating $b$ for an isothermal gas is fairly straightforward as the sound speed of isothermal gas is constant and identical everywhere, and $\mathcal{M}$ would just be the turbulent rms velocity dispersion ($\sigma_v$) divided by that global sound speed. This condition is satisfied for time $t = 0 \, \mathrm{kyr}$ in our simulation where all the gas is at $T \sim 10 \, \mathrm{K}$, and we obtain a value of $b \sim 0.43$; which is the value expected for isothermal gas initialised with turbulent velocity fluctuations having a natural mixture of compressive and solenoidal modes \citep{Federrath_2010}.

However, for $t > 0$ the situation is more complicated as there are a range of local sound speeds due to the ionizing radiation, and thus a range of local Mach numbers ($M$). Thus, we have to distinguish the rms Mach number $\mathcal{M}$ from the local Mach number $M$. Strictly speaking, Equation~(\ref{eq:b_parameter}) is only valid for an isothermal gas, as it has been derived and tested under isothermal conditions, and we can thus only calculate $b$ for the cold neutral gas in our simulations, i.e., by selecting regions of gas belonging to the cold neutral phase, and calculating the value of $b$ for them.

\subsubsection{Determining $\sigma_s$ and $\mathcal{M}$ for the dense, cold gas}
\label{sec:Neutral_Gas}
In this study we are concerned with the turbulence in the star-forming neutral gas that is driven by the surrounding ionized gas when irradiated with ionizing radiation. Hence we first pick out regions in our computational volume that have ionization fractions corresponding to neutral gas, i.e., $\eta < 10^{-7}$, which effectively corresponds to gas with temperature equal to the neutral gas temperature (i.e., $T = 10 \, \mathrm{K}$).

\begin{figure}
\centering
\includegraphics[width=\columnwidth]{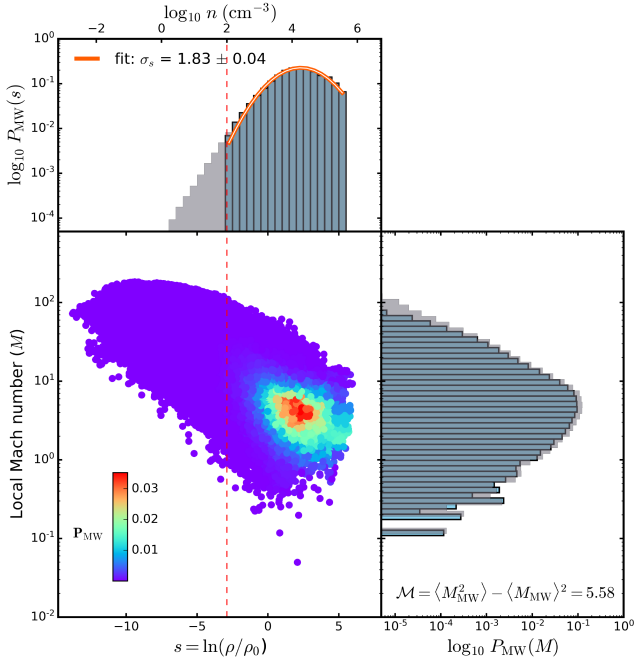}
\caption{Scatter plot of local Mach numbers ($M$) and scaled logarithmic densities, $s = \ln(\rho/ \rho_0)$, for the neutral gas in the entire simulation domain at time $t = 800 \, \mathrm{kyr}$, colour-coded by the mass-weighted probability density. Flanking histograms show the respective individual mass-weighted PDFs (grey) and the same for $n>10^2 \, \mathrm{cm}^{-3}$ (shaded blue). The derived value of $\sigma_s$ from a lognormal fit (Eq.~\ref{eq:lk03}, shown in orange) and $\mathcal{M}$ from the mass-weighted standard deviation of $M$ (computed from Eq.~\ref{eq:M_calculation}) are denoted in the respective panels. The values of $\sigma_s$ and $\mathcal{M}$ obtained with this method are then used in Equation~\ref{eq:b_parameter} to compute the driving parameter $b$.}
\label{fig:Global_Scatter}
\end{figure}

In Figure~\ref{fig:Global_Scatter} we show a scatter plot for $t = 800 \, \mathrm{kyr}$ of the local values of the scaled logarithmic density ($s$) and the Mach number ($M$), with the colourbar representing the mass-weighted probability density. We notice that over the wide range of $s$ and $Mach$ covered in the simulation, there is an anti-correlation between $s$ and $M$, which is a result of the fact the gas is not globally isothermal \citep{Federrath_2015}. Moreover, the geometry of the pillar-like structures, with a high positive gradient in the density away from the ionized gas towards the confined dense columns, results in the lower-density edges of the pillars having higher velocities as compared to the denser interiors, leading to the anti-correlation. In contrast, Equation~(\ref{eq:b_parameter}) only holds for the dense, cold, near-isothermal gas, where $s$ and $M$ are un-correlated \citep{Passot_1998,Kritsuk_2007,Federrath_2010}. 

In addition to this we notice that some of these low-density regions have extremely high local Mach numbers, with some of them exceeding $M \gtrsim 100$. Careful analysis of the data reveals that they lie at the shadows of the dense pillar tips, at the transition between the ionized and neutral phases of the gas, and the high velocities are the result of the large pressure gradient. However, as mentioned earlier, the cool temperatures of these regions ($T \sim 10 \, \mathrm{K}$) are numerical artefacts due to the uni-directional ray-tracer used and the lack of a diffuse EUV recombination field and would in reality have higher ionization fractions (and hence higher $c_s$), which would effectively reduce their $M$ to lower and more realistic values. To avoid the bias introduced by these artefacts, we 1) apply a density threshold of $n>n_\mathrm{threshold}=10^2\,\mathrm{cm}^{-3}$ and 2) use mass-weighted PDFs, such that the low-density regions with unrealistic values of $M$ do not contribute significantly to the density and velocity moments that enter Eq.~(\ref{eq:b_parameter}). This method allows us to measure the turbulence driving parameter $b$ for the dense, cold phase of neutral gas, where star formation would occur.

The flanking histograms in Figure~\ref{fig:Global_Scatter} show the mass-weighted PDFs for $s$ (integrated over $M$) and $M$ (integrated over $s$). The density threshold $n > n_\mathrm{threshold} = 10^2\,\mathrm{cm}^{-3}$ is indicated as a vertical dotted line, where $n_\mathrm{threshold}$ is the number density of the transition from atomic to molecular gas via surface reactions on dust grains. This value has been studied to lie in the range $100-1000 \, \mathrm{cm}^{-3}$ \citep{Glover_2007,Glover_2010}. Here we simply use $n_\mathrm{threshold} = 10^2 \, \mathrm{cm}^{-3}$, however, we also repeat our analyses for $n_\mathrm{threshold} = 1000 \, \mathrm{cm}^{-3}$ in Appendix~\ref{sec:Density_Threshold} and show that our results do not change significantly with the chosen density threshold. The shaded flanking PDFs overlaid on the overall neutral gas PDFs  in Figure~\ref{fig:Global_Scatter} show the distributions with the density threshold applied.

We now derive values for $\sigma_s$ and $\mathcal{M}$ from the distributions of $s$ and $M$, respectively. For the distribution of $s$ we fit the mass-weighted version of a lognormal given by \citet{Li_Klessen_Low_2003},
\begin{equation} \label{eq:lk03}
    P_\mathrm{LN}(s)\,ds = \frac{\mathrm{C} \exp(s)}{\sqrt{2 \pi \sigma_{s,\mathrm{MW}}^2}} \exp{
    \left[ -\frac{s-s_0+\sigma_{s,\mathrm{MW}}^2/2}{2\sigma_{s,\mathrm{MW}}^2} \right]}\,ds,
\end{equation}
where $s_0$ is mean value of $s$, $\sigma_{s,\mathrm{MW}}$ is the mass-weighted standard deviation in $s$, and $\mathrm{C}$ is a normalisation constant. We free all three parameters in this relation and compute their best fit values using a least-squares approach. We then set $\sigma_s$ to be equal to the best fit value of $\sigma_{s,\mathrm{MW}}$. To derive $\mathcal{M}$, we explicitly calculate the mass-weighted standard deviation of the local Mach number distribution given by
\begin{equation}
    \mathcal{M} = \sqrt{\langle M_{\mathrm{MW}}^2 \rangle - \langle M_{\mathrm{MW}} \rangle^2},
    \label{eq:M_calculation}
\end{equation}
where $\langle M_{\mathrm{MW}}^2 \rangle$ and $\langle M_{\mathrm{MW}} \rangle$ are the mean squared and arithmetic mean of the mass-weighted $M$, respectively. Figure~\ref{fig:Global_Scatter} demonstrates this method of analysis for the time snapshot $t = 800 \, \mathrm{kyr}$, with the derived values of $\sigma_s$ from the lognormal fit and $\mathcal{M}$ for the dense, cold gas denoted on the flanking histograms. This method is repeated for each time snapshot to obtain the time evolution of $b$.

\subsubsection{Time evolution of the global driving parameter $b$}

\begin{figure*}
\centering
\includegraphics[width=\textwidth]{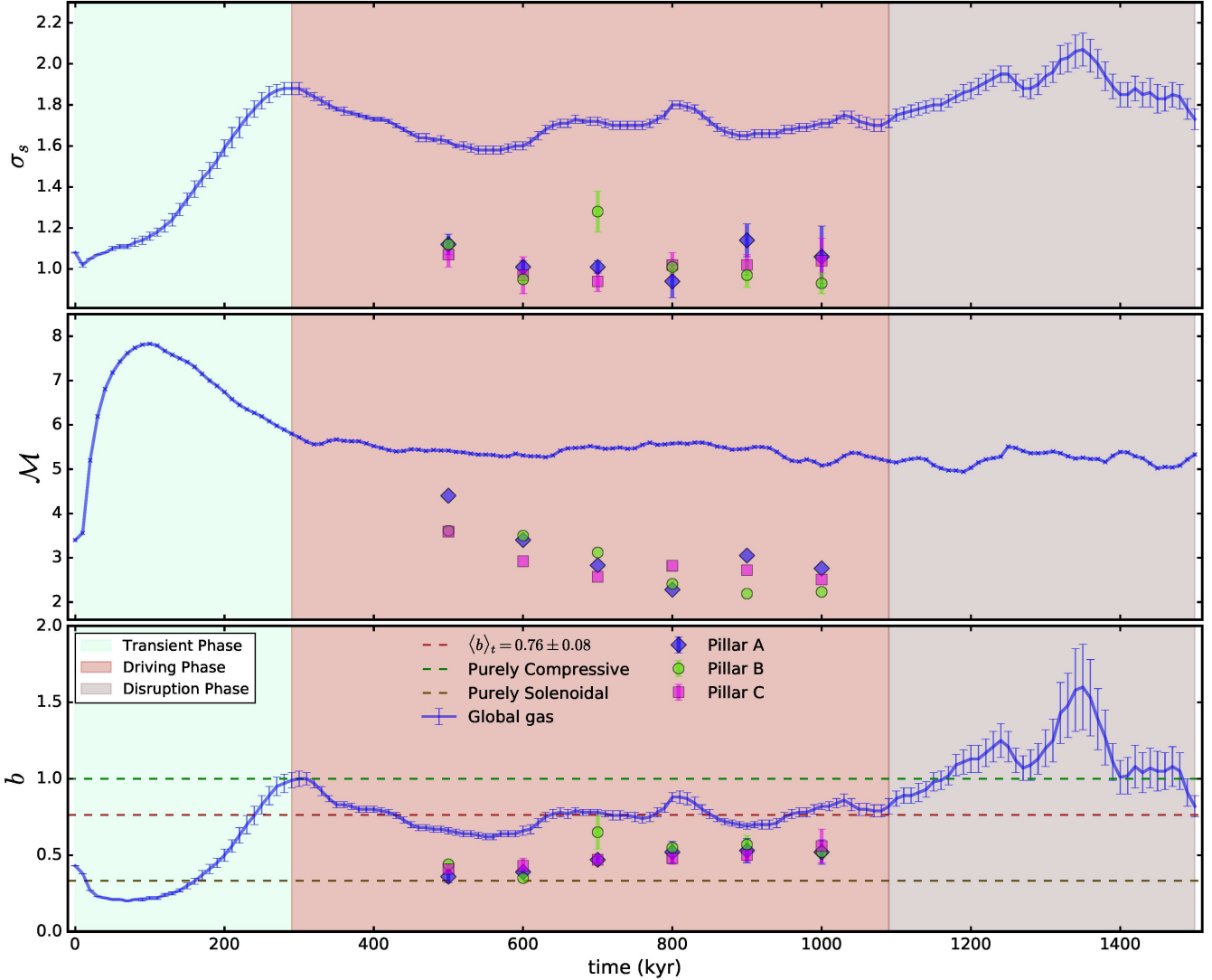}
\caption{Evolution of the driving parameter $b$ and the quantities involved in its calculation with time, where $t=0$ is the time when the ionization front hits the gas. Shaded regions indicate the phase of evolution as categorised in Section~\ref{sec:Neutral_Gas}. From top to bottom: logarithmic scaled density dispersion $\sigma_s$, rms Mach number $\mathcal{M}$, and driving parameter $b$. The dotted line in red indicates the time-averaged value of $b$ in the Driving Phase $\langle b \rangle_t=0.76\pm0.08$, and the values of $b$ corresponding to purely compressive (green) and purely solenoidal (violet) modes are also provided for reference, as dotted lines. The values of the same quantities obtained for the defined pillar regions (c.f.~Fig.~\ref{fig:Column_Density}), A (diamonds), B (circles), and C (squares), are shown for 5 different time snapshots, where these could be reasonably defined and followed in their individual evolution.}
\label{fig:b_timevariation}
\end{figure*}

Figure~\ref{fig:b_timevariation} shows the time evolution of $b$ with the method outlined in the previous subsection. We divide this time evolution into three broad temporal phases:

\begin{itemize}
    \item Transient Phase ($0\leq t<300\,\mathrm{kyr}$): the gas ionized by the radiation field is in the process of enhancing the initial overdensities seeded by the turbulence and sculpting pillar-like regions. The turbulence driven by the thermal pressure-induced compression of neutral gas, as evident from a sharply increasing value of $\mathcal{M}$, pushes neutral gas from lower density columns into denser ones, resulting in an enhancement of the initial overdensities and increasing $\sigma_s$. We observe a time lag in the increase of $\sigma_s$ and $\mathcal{M}$, as one is the cause (higher $\mathcal{M}$) and the other the effect (higher $\sigma_s$), and this lag biases the value of $b$ to be lower at very early times (up to $\sim 200\,\mathrm{kyr}$). When the velocity fluctuations start to succeed in enhancing the initial overdensities by dissipating their energy in shocks, $\mathcal{M}$ starts to decrease until it reaches a steady value of $\mathcal{M}\sim5.5$. 
    
    \item Driving Phase ($300\leq t<1100\,\mathrm{kyr}$): pillar-like structures have already been created as a result of the ionizing radiation-driven compression and the dynamics of the neutral gas have been established, albeit subject to local structural changes due to compression or photo-evaporation at the individual pillar regions (as seen from the temporal fluctuations in $\sigma_s$). We thus call this phase the 'Driving Phase‘, as the turbulence has settled down and thus a driving parameter $b$ can be reasonably defined here. As we can see in Figure~\ref{fig:b_timevariation}, a fairly stationary value of $b$ is obtained in this phase. We calculate the time-averaged value of $b$ in this phase as $\langle b \rangle_t = 0.76 \pm 0.08$. 
    
    \item Disruption Phase ($1100\leq t<1500\,\mathrm{kyr}$): at these late times of the evolution, many of the neutral pillar-like structures get either expelled through the outflow boundaries of the computational box and/or get successfully ionized by the expanding HII region. The amount of neutral molecular gas available in the cloud is insufficient to derive meaningful statistics, as is evident from the relatively large error bars in this final phase of evolution. 
\end{itemize}

We thus argue that the value of $b$ obtained in what we call the 'Driving Phase‘ most sensibly represents what could be characterised as the turbulence driving mode of ionizing radiation. The time-averaged $b \sim 0.76$ suggests that the turbulence driven in the neutral molecular gas by the ionizing radiation on cloud scales ($\sim 4 \; \mathrm{pc}$) is predominantly compressive ($b>0.4$) in nature.

\begin{figure}
\centering
\includegraphics[width=\columnwidth]{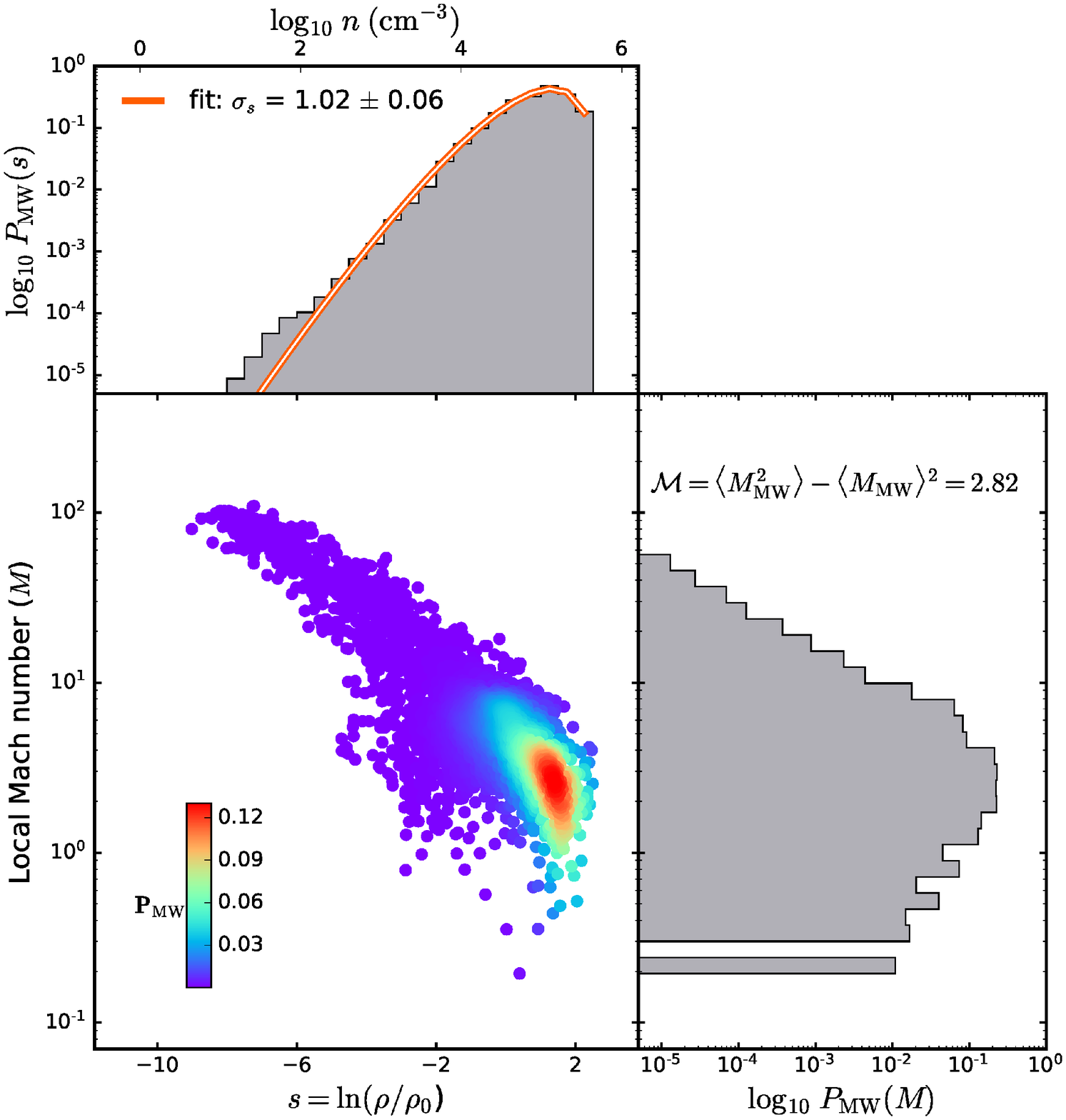}
\caption{Same as Figure~\ref{fig:Global_Scatter}, but for the sub-parsec region Pillar C at $t = 800 \, \mathrm{kyr}$, demonstrating our analysis method to derive values of $b$ for the three pillar regions. We do not use a density cutoff here as the contributions from $n<100 \; \mathrm{cm}^{-3}$ are negligible. We use the \citet{Hopkins_2013} intermittency PDF model (Eq.~\ref{eq:Hopkins}) instead of a standard lognormal to fit for $\sigma_s$ (shown in orange). }
\label{fig:Pillar_Scatter}
\end{figure}

\subsection{Pillar Regions}

Studies show that star formation is observed at the tips of pillar-like structures in a molecular cloud irradiated by ionizing radiation. Information on the local dynamics of the region is thus important to understand the star formation potential of the gas. Hence we now calculate the driving parameter $b$ separately for the three sub-parsec-scale pillar regions denoted in Figure~\ref{fig:Column_Density}, namely Pillar A, B and C. The regions all have cubical sizes of $(0.4 \, \mathrm{pc})^3$, adjusted such that the centre of mass of the region lies at the centre of the defined cubical box. We then select again only the neutral gas in these regions ($\eta < 10^{-7}$). However, we do not set any density threshold for pillar regions as the contribution of gas with $n<n_\mathrm{threshold}$ is negligible. We derive the Mach number ($\mathcal{M}$) as we did for the global gas, i.e., as the standard deviation of the mass-weighted distribution of the local Mach number ($M$), and given by Equation~\ref{eq:M_calculation}. However, the distribution of $s$ is more skewed, and a symmetrical lognormal does not fit the data well on the local pillar scales, most likely because of intermittency in the turbulence \citep{Kritsuk_2007,Federrath_2010,Hopkins_2013,Squire_2017}. To account for this skewness, we instead fit the mass-weighted version of the \cite{Hopkins_2013} intermittency PDF model,
\begin{multline}
\label{eq:Hopkins}
    p_{\mathrm{HK}}(s)\,ds = \mathrm{C} e^s I_1 (2 \sqrt{\lambda \omega(s)}) \exp {\left[ -(\lambda + \omega(s)) \right]} \sqrt{\frac{\lambda}{\theta^2 \omega(s)}}\,ds, \\
    \lambda = \frac{\sigma_{s,\mathrm{MW}}^2 (1+\theta)^3}{2 \theta^2}, \quad \omega(s) = \lambda/(1+\theta) - s/\theta \; (\omega \geq 0),
\end{multline}

where $I_1(x)$ is the first-order modified Bessel Function of the first kind, $\sigma_{s,\mathrm{MW}}$ is the mass-weighted standard deviation in $s$, and $\theta$ is the intermittency parameter. Note that in the zero-intermittency limit ($\theta \to 0$) Eq.~(\ref{eq:Hopkins}) simplifies to the lognormal PDF. We fit Equation~\ref{eq:Hopkins} to $P_{\mathrm{MW}}(s)$ and determine the best-fit $\sigma_{s,\mathrm{MW}}$. We show the scatter of $s$ and $M$, their corresponding mass-weighted histograms, and the values of $\sigma_{s}$ and $\mathcal{M}$ in Figure~\ref{fig:Pillar_Scatter} for time snapshot $t = 800 \, \mathrm{kyr}$. These are then used in Equation~\ref{eq:b_parameter} to obtain the value of $b$. This is repeated for five different time snapshots in the Driving Phase and the obtained values were added in Figure~\ref{fig:b_timevariation}.

Figure~\ref{fig:b_timevariation} shows that the values for $\sigma_s$ and $\mathcal{M}$ for the pillar regions are lower than that for the gas on the global cloud scales. This is expected from the observed cloud scaling relations, such as the velocity dispersion--size relation. The pillar regions are considerably smaller ($\sim0.4\,\mathrm{pc}$) and thus have only mildly supersonic velocity dispersions (\mbox{$\mathcal{M}\sim2$--$4$}). The obtained velocity dispersions ($\mathcal{M}\times c_\mathrm{s}$) are in the range \mbox{$0.4$--$1\,\mathrm{km}\,\mathrm{s}^{-1}$}, in agreement with observations of pillars on these scales \citep{Klaassen_2019}.

\begin{figure}
\centering
\includegraphics[width=\columnwidth]{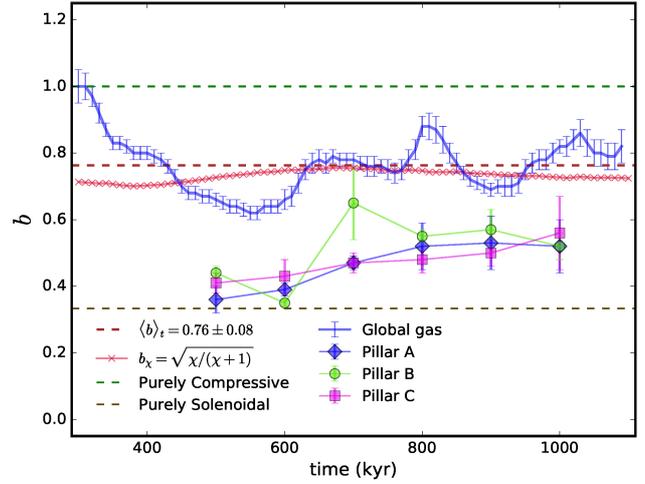}
\caption{Zoom-in of the time evolution of the driving parameter $b$ during the Driving Phase for the three pillar regions. Dotted lines indicate purely compressive (green) and purely solenoidal (brown) values of $b$ for reference. Solid lines indicate the time evolution of $b$ for the gas on $\sim 4 \, \mathrm{pc}$ global cloud scales (blue) and the value of $b$ inferred from the compressive ratio $b_{\chi}$ (red) discussed in Section~\ref{sec:compratio}.}
\label{fig:Pillar_bvalues}
\end{figure}

The $b$ values of the pillar regions are also somewhat smaller with \mbox{$b\sim0.4$--$0.6$} compared to the cloud-scale average. Interestingly, we find a noticeable time variation for each pillar region, individually. This is more clearly seen in Figure~\ref{fig:Pillar_bvalues}, where we find that the value of $b$ increases for each pillar, from a value of $b \sim 0.4$ to $0.55$ as time progresses, indicating that the turbulence in these regions develops more compressive modes as they evolve. This might suggest that conditions for star formation are progressively boosted in these regions, as the HII region moves through the cloud. That being said, commenting further on the potential star formation in these simulations is outside the scope of this work as it would require follow-up simulations that include self-gravity and a model for star formation such as the sink-particle technique \citep[e.g.,][]{Federrath_2010_Sinks}. 

\subsection{The compressive ratio} \label{sec:compratio}

So far we have primarily used the density dispersion and rms Mach number to study the turbulence driving parameter $b$ of clouds hit by an ionization front. However, we can also look at the compressible-to-solenoidal mode mixture in the velocity field to obtain related information about $b$, via the so-called 'compressive ratio‘ \citep{Kritsuk_2007,Federrath_2010,Federrath_PRL,Pan_2016,Kortgen_2017}. The compressive ratio is given by 
\begin{equation}
\chi = \langle v_c^2 \rangle / \langle  v_s^2 \rangle,
\end{equation}
where $v_c$ and $v_s$ are the compressive and solenoidal components of the velocity field, respectively. In order to compute these two components, the velocity field is first corrected for bulk motion, such that only the turbulent fluctuations remain \citep[analogous to $\chi_\mathrm{turb}$ defined in][]{Pan_2016}, and then density-weighted to trace the velocity field of the cold, dense gas. Both $v_c$ and $v_s$ are then derived through a Helmholtz decomposition of the turbulent velocity field in Fourier space. Following the suggestion of \citet{Pan_2016}, an effective driving parameter can be derived from this value of $\chi$ given by $b_{\chi} = \sqrt{\chi/(\chi+1)}$.

The evolution of $b_{\chi}$ during the Driving Phase is shown in Figure~\ref{fig:Pillar_bvalues}. As we can see the value we obtain remains more or less constant ($\sim 0.75$) and indicates strong compressive velocity fields. This value is also very close to the time-averaged value $\langle b \rangle_t$ obtained earlier from the relation between the density and velocity fluctuations (Equation \ref{eq:b_parameter}). The fact that $b$ obtained from the \mbox{$\sigma_s$--$\mathcal{M}$} relation (Eq.~\ref{eq:sigmas}) agrees with that obtained from the modes in the velocity field ($b_{\chi}$) is encouraging and supports our overall conclusions of a relatively compressive turbulence driving parameter ($b>0.4$) for clouds compressed in HII regions.

\subsection{Virial Parameter} \label{sec:vir_parameter}

The virial parameter is a dimensionless quantity that characterises the ratio of turbulent kinetic energy to gravitational energy of a cloud of gas \citep{Bertoldi_1992}, with its general form given by \citep{federrath_klessen_2012},
\begin{equation}
	\alpha_{\mathrm{vir}} = 2E_{\mathrm{kin}}/|E_{\mathrm{grav}}|, 
\end{equation}
where $E_{\mathrm{kin}}$ and $E_{\mathrm{grav}}$ are the kinetic and potential energies of the cloud, respectively. A value of $\alpha_{\mathrm{vir}} < 1$ suggests that the cloud could be gravitationally unstable and thus potentially form stars, with the star formation rate per free fall time $\mathrm{SFR}_{\mathrm{ff}}$ increasing with decreasing $\alpha_{\mathrm{vir}}$ \citep{Krumholz_Mckee_2005,Hennebelle_Chabrier_2011,Padoan_2012,federrath_klessen_2012}.
 
\begin{figure}
\centering
\includegraphics[width=\columnwidth]{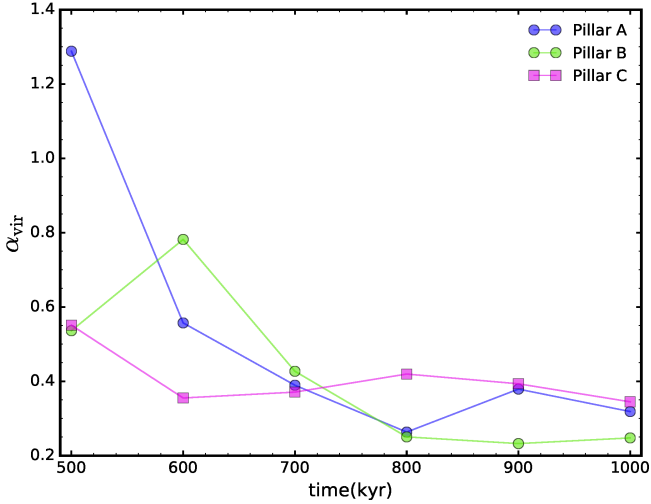}
\caption{Time evolution of the virial parameter $\alpha_{\mathrm{vir}}$ computed with Eq.~(\ref{eq:vir}) for each of the three pillar regions.}
\label{fig:Pillar_Virial}
\end{figure}

We therefore study the time evolution of $\alpha_{\mathrm{vir}}$ for our defined pillar regions to test whether the driving of compressive turbulence is accompanied by an increase in the star-formation capability of the gas. We calculate $\alpha_{\mathrm{vir}}$ by computing the self-gravitational potential $\Phi$ for each pillar region from their respective density distributions, with boundary conditions taking into account the density distribution of gas outside the boundaries of the pillar regions. This is important, because the binding energy of pillars is not independent of the environment \citep{federrath_klessen_2012}, as these regions are not isolated, but instead are typically deeply embedded inside a larger molecular cloud complex. We calculate $\alpha_{\mathrm{vir}}$ as 
 \begin{equation} \label{eq:vir}
 	\alpha_{\mathrm{vir}} = \frac{\sum_{i \in \mathrm{P}} m_i |(\mathbf{v}_{\mathrm{turb}})_i|^2}{\sum_{i \in \mathrm{P}} m_i |\Phi_i|} ,
 \end{equation}
where $\mathrm{P}$ is the set of computational cells lying in the pillar region, $m_i$ the mass, and $(\mathbf{v}_{\mathrm{turb}})_i$ the turbulent velocity in cell $i$. The time evolution of this calculated $\alpha_{\mathrm{vir}}$ for the pillars is shown in Figure~\ref{fig:Pillar_Virial}. 

As we can see the value of $\alpha_{\mathrm{vir}}$ goes down with time for all the pillar regions in our simulations. This is because the turbulent kinetic energy in the pillars is dissipated in shocks produced by the compressive turbulence (decreasing $|(\mathbf{v}_{\mathrm{turb}})_i|$ locally). This leads to the formation of local overdensities in the pillars (increasing $|\Phi_i|$ locally), which effectively results in lower values of $\alpha_{\mathrm{vir}}$. This suggests that the driving of compressive turbulence in the pillars lowers the value of the virial parameter and as a result, could potentially trigger star formation in them. We note that there are alternatives to Equation~(\ref{eq:vir}) to calculate the virial parameter, especially in observational studies where the self-gravitational potential is not available. In Appendix~\ref{sec:Virial_Parameter_Appendix} we test these different methods, and find that although there is a significant difference in the absolute values of $\alpha_{\mathrm{vir}}$ obtained with different methods, the trend of a decreasing $\alpha_{\mathrm{vir}}$ with time is independent of the choice of method.   

\section{Summary and conclusions}

In this study we simulate the formation of pillar-like structures in turbulent clouds at the edge of a Str{\"o}mgren sphere, by an expanding HII region modelled as an impinging plane-parallel ionizing front. We form structures that resemble observed pillars in morphology and dynamics, and study their evolution for up to $2\,$~Myr. We infer the effective mode of turbulence that is driven by the ionization front in the dense, cold gas of the cloud, by calculating the turbulence driving parameter ($b$) in Equation~(\ref{eq:sigmas}). We do the same for three different sub-parsec scale regions at the tip of pillars containing the densest gas, where star formation is often observed. Our main results can be summarised as follows:

\begin{enumerate}
    \item The value of $b$ for the cold, dense gas in the cloud does not vary significantly with time during the Driving Phase (\mbox{$300$--$1100\,\mathrm{kyr}$}), with a time-averaged value of $\langle b \rangle_t = 0.76 \pm 0.08$. This suggests that expanding HII regions drive predominately compressive modes of turbulence.
    \item The values of $b$ calculated for the pillar regions are also primarily compressive albeit with lower values of \mbox{$b\sim 0.4$--$0.6$} compared to the gas on global cloud scales. We also obtain values for the 3D velocity dispersion (\mbox{$\sigma_v\sim0.4$--$1\,\mathrm{km}\,\mathrm{s}^{-1}$}) that agree with observations in the pillar regions. 
    \item The turbulent modes for each of the pillar regions transition from a natural mixture of solenoidal and compressive modes ($b \sim 0.4$) to a compression-dominated regime ($b \sim 0.55$) during their lifetime, as the ionized gas continues to sculpt the pillars.
    \item We calculate the compressive ratio ($\chi$) for the cloud and infer a driving parameter $b_{\chi} \sim 0.75$ from it, consistent with the time-averaged global $b$ value obtained from the density dispersion--Mach number relation, supporting our main conclusion that expanding HII regions drive primarily compressive modes of turbulence.
    \item The virial parameter $\alpha_{\mathrm{vir}}$ decreases with time for the pillar regions, which suggests that the driving of compressive turbulence in these regions, as the HII region passes through the cloud, is accompanied by an increase in the star-formation capability of the gas.
\end{enumerate}

Our main finding of predominantly compressive turbulence driven in HII regions may be interpreted as promoting star formation, and hence leading to 'triggered star formation'. However, the converse effect of the neutral gas being photo-evaporated by the radiation limits the extent of this triggering, and the net star formation occurring in the region is likely the result of the competition between these two processes. Follow-up simulations with self-gravity and sink particles to follow star formation would allow us to make more quantitative predictions on the net effect of photo-ionization on star formation in molecular clouds.
\label{sec:summary}

\section*{Acknowledgements}
We thank Richard Wunsch for a timely and very constructive referee report. SHM and RK acknowledge financial support via the Emmy Noether Research Group on Accretion Flows and Feedback in Realistic Models of Massive Star Formation funded by the German Research Foundation (DFG) under grant no. KU 2849/3-1 and KU 2849/3-2. C.~F.~acknowledges funding provided by the Australian Research Council (Discovery Project DP170100603 and Future Fellowship FT180100495), and the Australia-Germany Joint Research Cooperation Scheme (UA-DAAD). We acknowledge support by the High Performance and Cloud Computing Group at the Zentrum für Datenverarbeitung of the University of Tübingen, the state of Baden-Württemberg through bwHPC and the German Research Foundation (DFG) through grant no INST 37/935- 1 FUGG. We further acknowledge high-performance computing resources provided by the Leibniz Rechenzentrum and the Gauss Centre for Supercomputing (grants~pr32lo, pr48pi and GCS Large-scale project~10391), the Australian National Computational Infrastructure (grant~ek9) in the framework of the National Computational Merit Allocation Scheme and the ANU Merit Allocation Scheme.




\bibliographystyle{mnras}
\bibliography{iondriving} 




\appendix
\section{Resolution Study}
\label{sec:Resolution_Study}
\begin{figure*}
    \centering
    \includegraphics[width=\columnwidth]{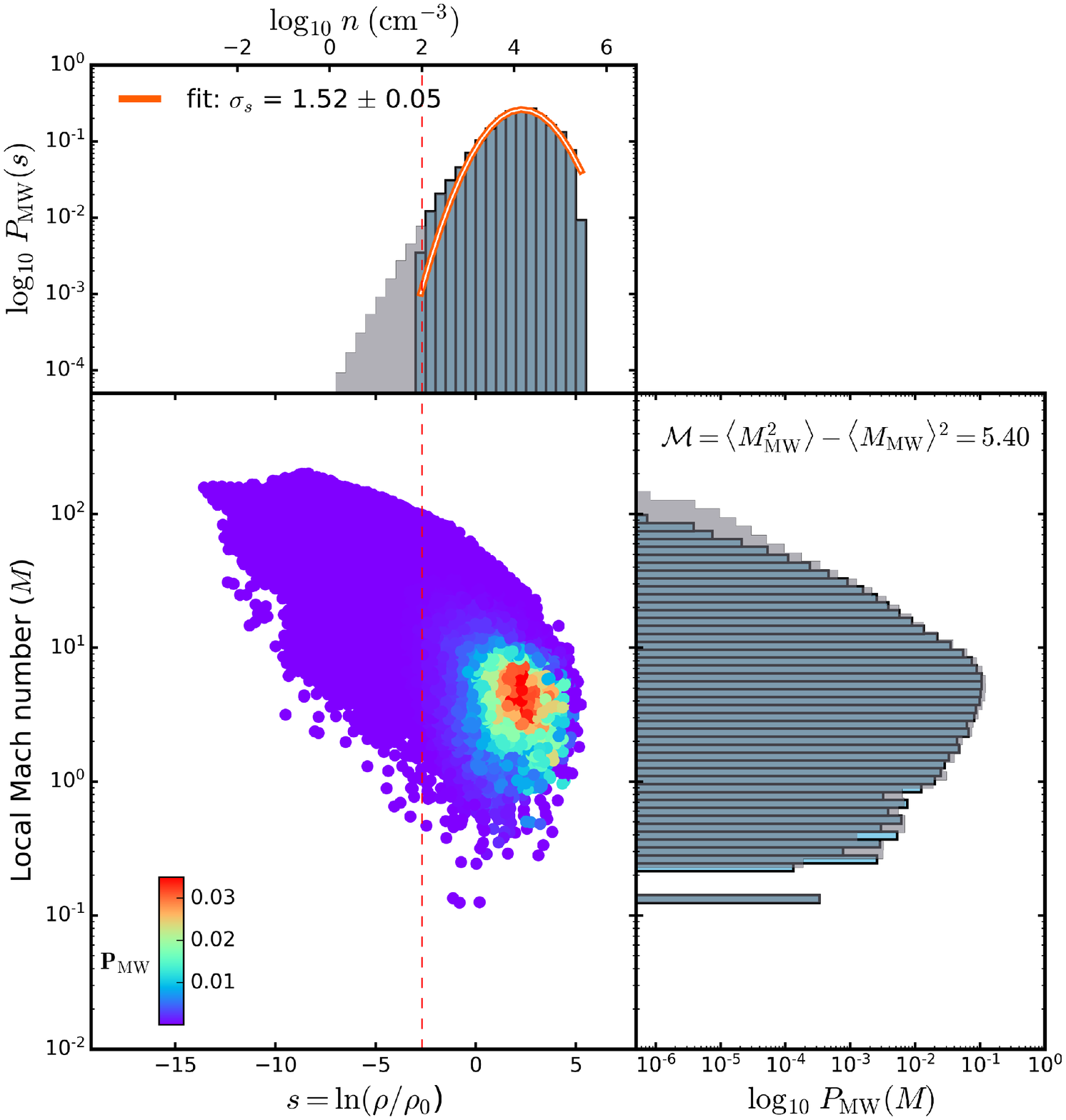}
    \includegraphics[width=\columnwidth]{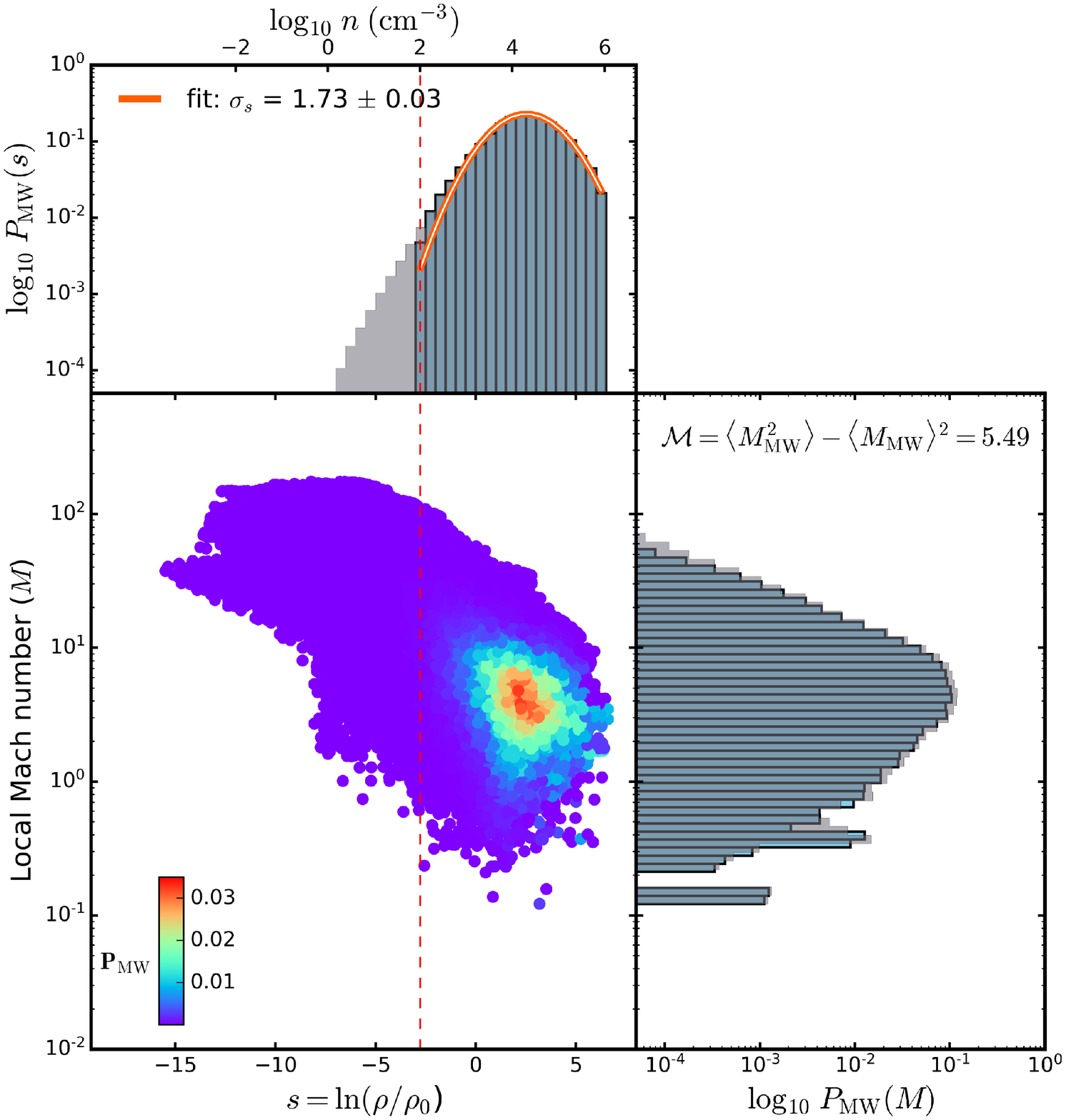}
    \includegraphics[width=\columnwidth]{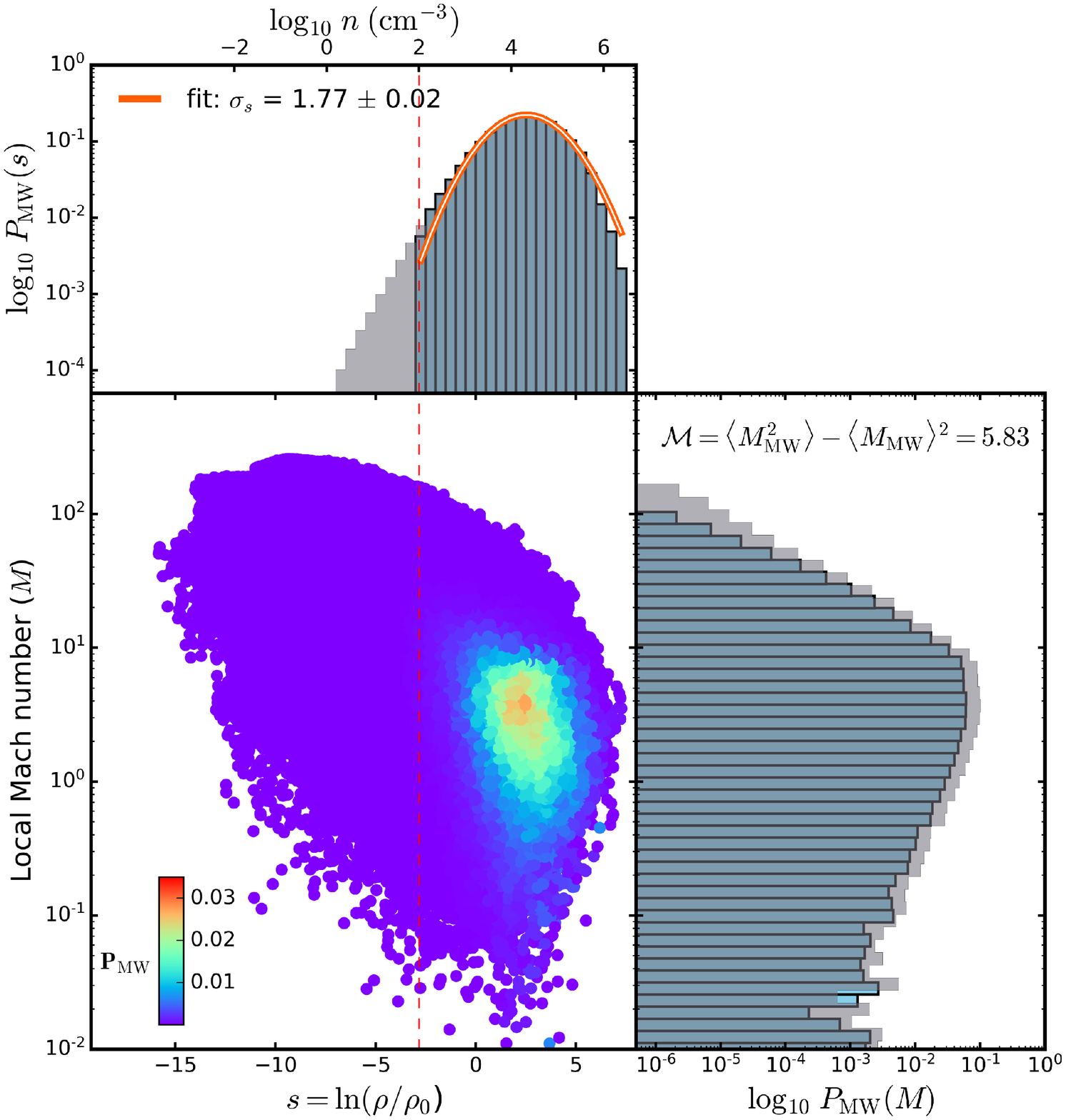}
    \caption{Scatter plot and flanking histograms (similar to Figure~\ref{fig:Global_Scatter}) for $t = 700 \, \mathrm{kyr}$ comparing three different numerical grid resolutions. Top-left : $100^3$ ($\Delta x = 0.04 \, \mathrm{pc}$), top-right : $200^3$ ($\Delta x = 0.02 \, \mathrm{pc}$) and bottom:  $400^3$ ($\Delta x = 0.01 \, \mathrm{pc}$).}
    \label{fig:Resolution_Scatter}
\end{figure*}

\begin{figure*}

\centering
\includegraphics[width=\textwidth]{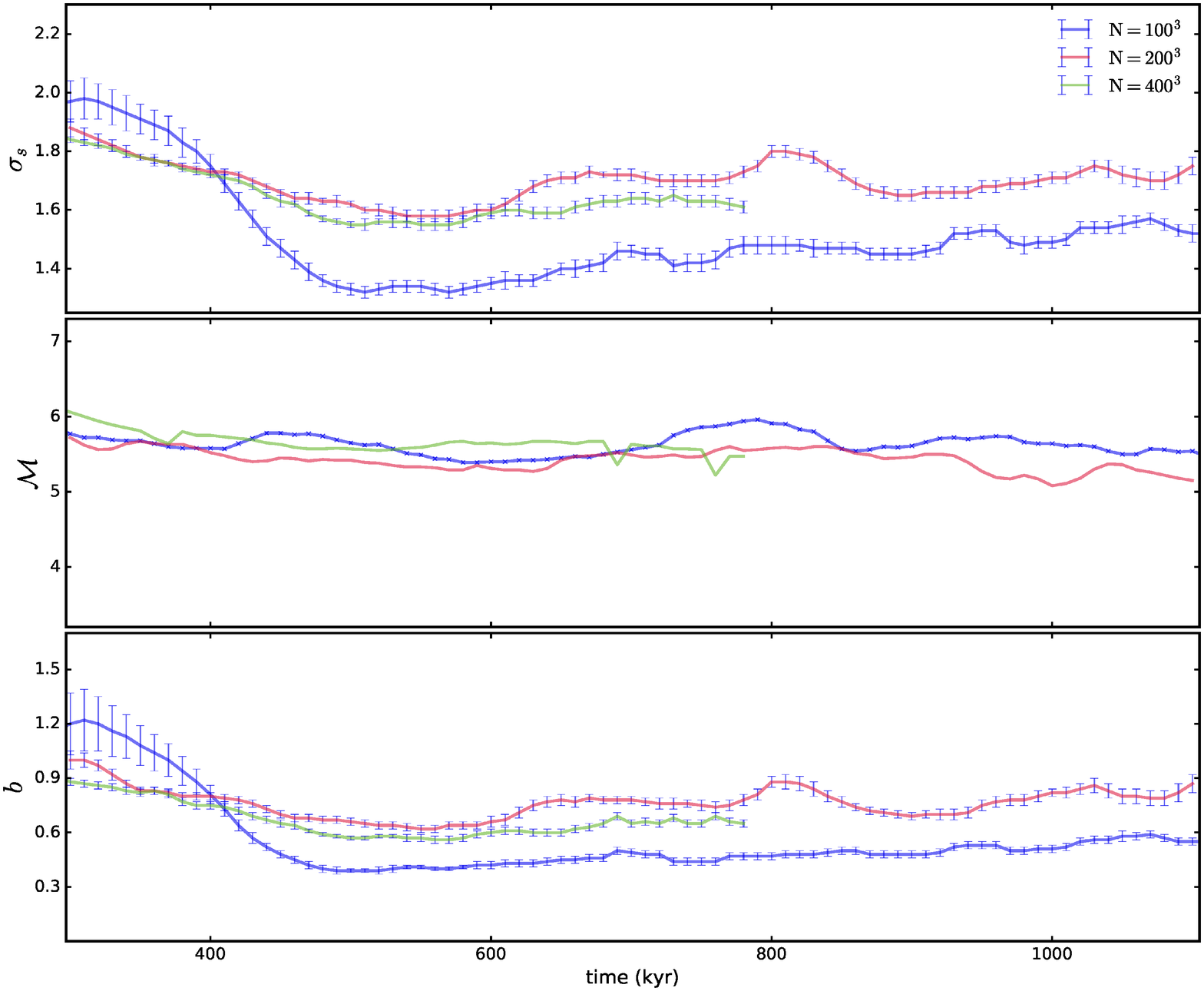}

\caption{The values of $\sigma_s$, $\mathcal{M}$, and $b$ compared for simulations with resolutions of $100^3$ ($\Delta x = 0.04 \, \mathrm{pc}$), $200^3$ ($\Delta x = 0.02 \, \mathrm{pc}$) and $400^3$ ($\Delta x = 0.01 \, \mathrm{pc}$). We find reasonable convergence for resolutions of $\gtrsim200^3$ grid cells ($\Delta x \lesssim 0.02 \, \mathrm{pc}$).}
\label{fig:Resolution_Convergence}
\end{figure*}

The decay rate of the turbulent energy and the general properties of turbulence in numerical simulations depend on the numerical resolution adopted, with lower resolution showing, e.g., a larger (artificial) decay \citep{Mac_Low_1998,Federrath_2013}. In order to test numerical convergence of our results, we compare $\sigma_s$, $\mathcal{M}$ and the resulting $b$ parameter (from Eq.~\ref{eq:b_parameter}) for three different numerical grid resolutions: $100^3$ ($\Delta x = 0.04 \, \mathrm{pc}$), $200^3$ ($\Delta x = 0.02 \, \mathrm{pc}$), and $400^3$ ($\Delta x = 0.01 \, \mathrm{pc}$). We note that the $400^3$ simulation has only been followed for comparison purposes up to $\sim 800 \, \mathrm{kyr}$ due to the computational cost of that simulation. Figure~\ref{fig:Resolution_Scatter} shows the scatter plot and corresponding PDFs of $s$ and $M$ for the three different resolutions at $t = 700 \, \mathrm{kyr}$, and Figure~\ref{fig:Resolution_Convergence} compares the time evolution of $\sigma_s$, $\mathcal{M}$ and $b$ derived from the PDFs, following the method explained in the main part of the manuscript. We can see that $\mathcal{M}$ is almost independent of the choice of resolution (as long as the grid resolution is at least $100^3$). However, $\sigma_s$ and $b$ show a systematic increase between $100^3$ and $200^3$, while for $\gtrsim 200^3$, the resulting $\sigma_s$ and $b$ are close to the values obtained for $200^3$ grid cells. Similar trends for the dependence of $\mathcal{M}$ and $\sigma_s$ are seen in \citet{Federrath_2010}, \citet{Kitsionas_2009}, and later resolution studies. We thus conclude that a spatial resolutions of $\lesssim0.02 \, \mathrm{pc}$ is sufficient to achieve reasonable convergence in this type of simulation.

\section{Choice of density threshold}
\label{sec:Density_Threshold}

\begin{figure*}
\centering
\includegraphics[width=\textwidth]{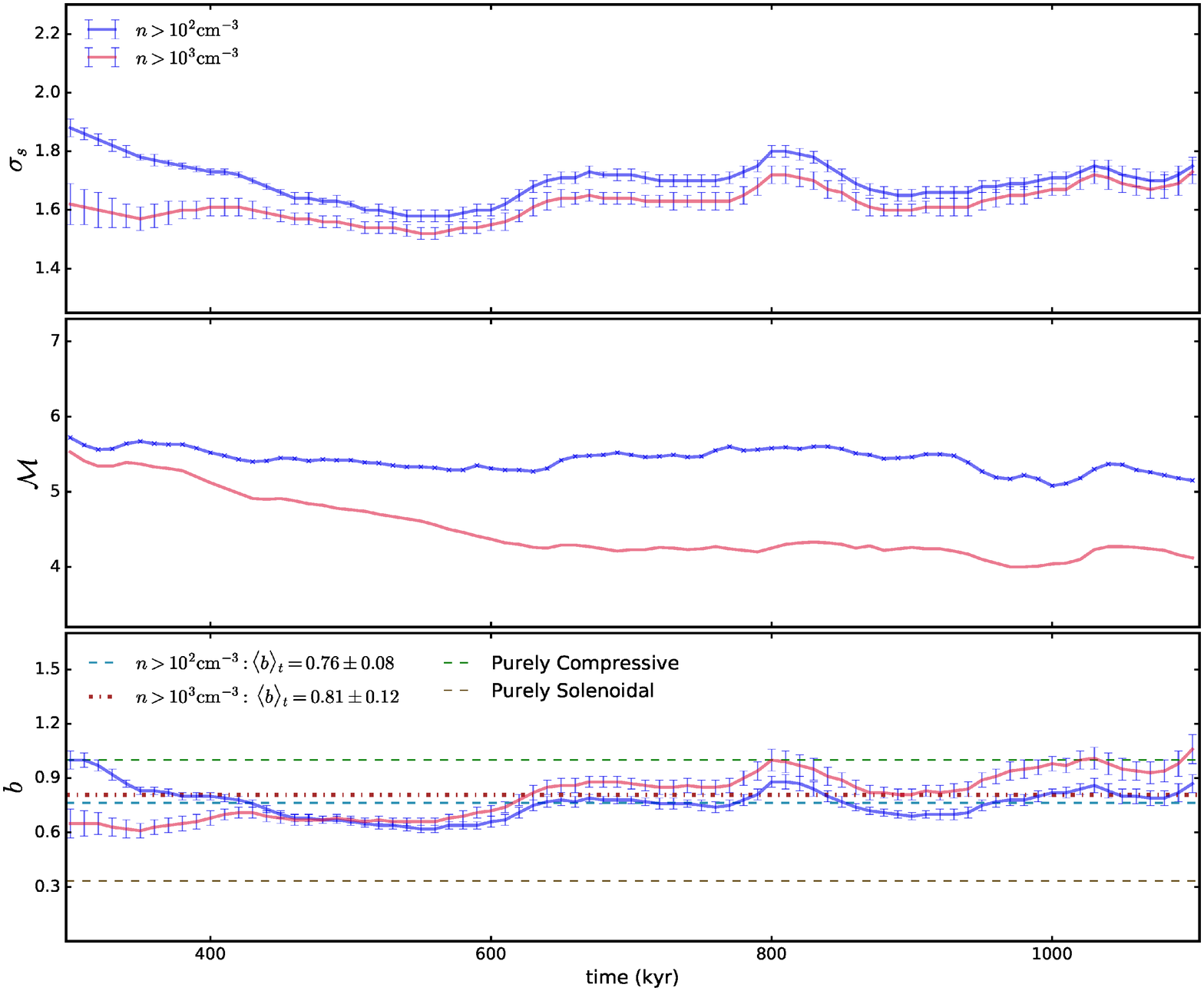}
\caption{Comparison of derived values of $\sigma_s$, $\mathcal{M}$, and $b$ with $n_\mathrm{threshold} = 100 \, \mathrm{cm}^{-3}$ and $1000 \, \mathrm{cm}^{-3}$. We find that our results are not particularly sensitive to the choice of $n_\mathrm{threshold}$ as long as $n_\mathrm{threshold}$ is in the reasonable range \mbox{$10^2$--$10^3 \, \mathrm{cm}^{-3}$} to define dense, cold gas.}
\label{fig:Density_Threshold}
\end{figure*}

In this study we use a threshold for the number density, $n_\mathrm{threshold}=10^2\,\mathrm{cm}^{-3}$, for selecting dense, cold gas, representing the minimum number density where the transition from atomic to molecular gas via surface reactions on dust grains occurs. This value is expected to be in the range $100 - 1000 \; \mathrm{cm}^{-3}$ as shown in e.g., \cite{Glover_2007}, and thus we adopt a fiducial value of $n_\mathrm{threshold} = 10^2 \, \mathrm{cm}^{-3}$ for our main analyses. However, here we show that our results do not significantly depend on the exact choice of $n_\mathrm{threshold}$, by comparing results for $n_\mathrm{threshold} = 10^2 \, \mathrm{cm}^{-3}$ to $10^3 \, \mathrm{cm}^{-3}$ in Figure~\ref{fig:Density_Threshold}. Apart from the initial transient phase (where the amount of gas satisfying the condition $n > 1000 \, \mathrm{cm}^{-3}$ is very low and hence the error bars are very high), the values of $b$ for both density thresholds agree to within $<10\%$ deviation. The time-averaged values of $b$ in the Driving Phase are $0.76\pm0.08$ and $0.81\pm0.12$ for $n_\mathrm{threshold} = 10^2 \, \mathrm{cm}^{-3}$ and $10^3\,\mathrm{cm}^{-3}$, respectively. We can thus conclude that adopting any value for $n_\mathrm{threshold}$ in the realistic range of the transition number density (\mbox{$100$--$1000\,\mathrm{cm}^{-3}$}) gives robust results.
 
\section{Virial parameter method}
\label{sec:Virial_Parameter_Appendix}

In Section~\ref{sec:vir_parameter} we compute the virial parameter ($\alpha_{\mathrm{vir}}$) for the pillars by calculating the ratio of their kinetic and potential energies, which are individually estimated as the sum of the local contributions to the energies by computational cells belonging to the region of interest. However, there are alternative methods used to calculate $\alpha_{\mathrm{vir}}$, and the values obtained may depend on the method. To test the dependence of $\alpha_\mathrm{vir}$ on the method, we explore two alternative methods to calculate the virial parameter, namely

\begin{itemize}
	\item using a fixed velocity dispersion. In this, case we do not take into account the individual contributions of cells in the calculation of the kinetic energies, but rather estimate it from the velocity dispersion ($\sigma_v$) for the entire pillar region. $\alpha_{\mathrm{vir}}$ is then
	\begin{equation}
		\alpha_{\mathrm{vir}} = \frac{\sigma_v ^2\sum_{i \in \mathrm{P}} m_i }{\sum_{i \in \mathrm{P}} m_i |\Phi_i|} ,
		\label{eq:vp1}
	\end{equation}
	\item using an approximation that treats the pillar as if it were an isolated, spherical region. For this method, we assume the pillars can be approximated as spherical clouds of mass $M_c = \sum_{i \in \mathrm{P}} m_i$ and radius $R_c$ occupying a volume $V = 4/3 \pi R_c^3 = \sum_{i \in \mathrm{P}} V_i$, where $V_i$ is the volume occupied by computational cell $i$. $\alpha_{\mathrm{vir}}$ is then estimated as 
	\begin{equation}
		\alpha_{\mathrm{vir}} = \frac{5 \sigma_v^2L}{6GM_c}, 
		\label{eq:vp2}
	\end{equation}
	where $L = 2R_c$, $G$ is the gravitational constant, and $\sigma_v$ is the velocity dispersion in the pillar. This is the method widely used in observational studies to determine $\alpha_{\mathrm{vir}}$, because the 3D gravitational potential is not available in observations.
\end{itemize}	

\begin{figure}
\centering
\includegraphics[width=\columnwidth]{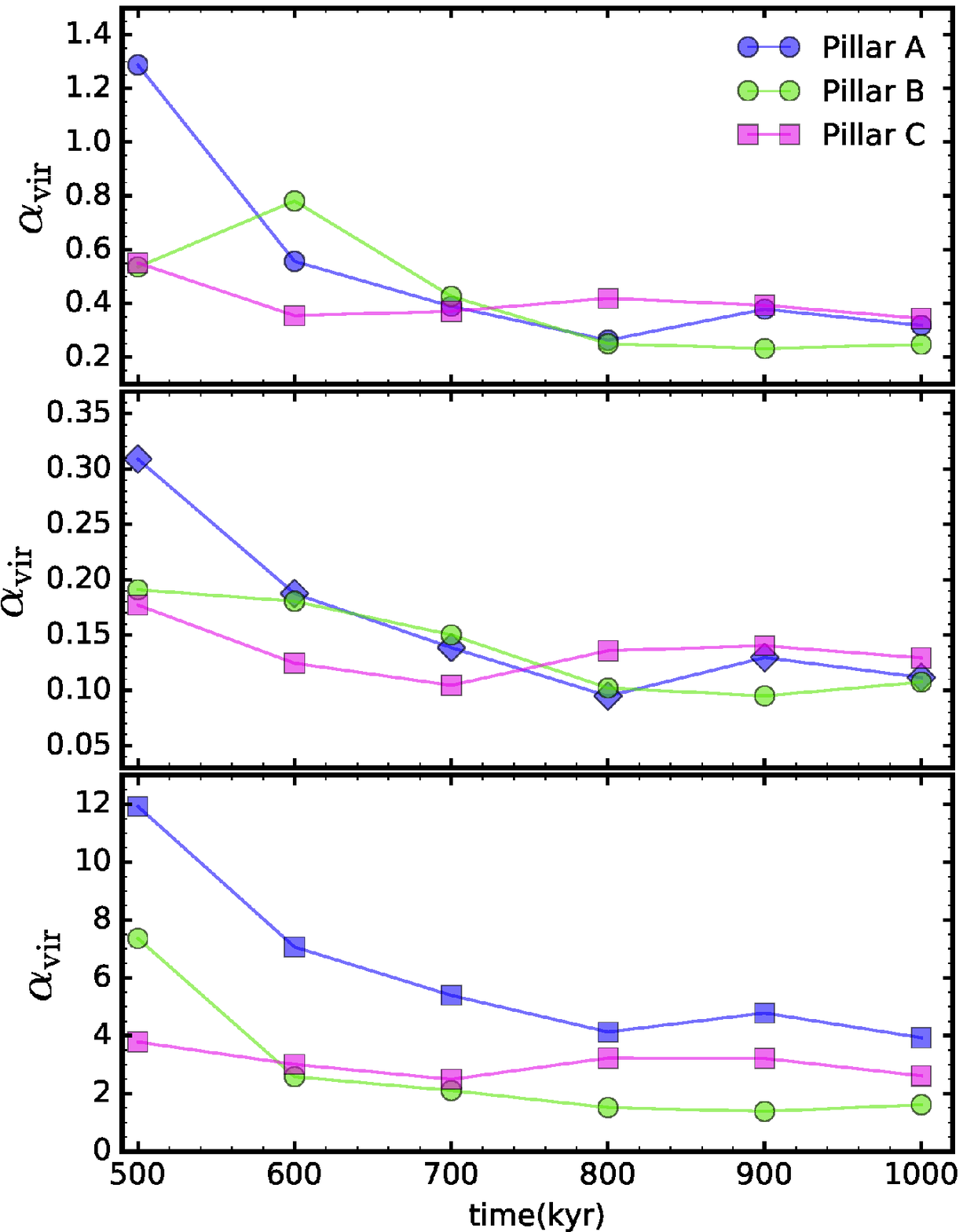}
\caption{Evolution of the virial parameter $\alpha_{\mathrm{vir}}$ with time obtained for the pillar regions using three different methods: 1) the method defined in Section~\ref{sec:vir_parameter}, which uses all local information to compute $\alpha_\mathrm{vir}$ (top panel), 2) using a fixed velocity dispersion (middle panel), given by Equation~(\ref{eq:vp1}), and 3) using the isolated, spherical-cloud approximation (bottom panel), given by Equation~(\ref{eq:vp2}). We find that our result of a decreasing trend for $\alpha_{\mathrm{vir}}$ with time is robust, however, the value of $\alpha_{\mathrm{vir}}$ changes significantly with the choice of method.}
\label{fig:vir_comparison}
\end{figure}

In Figure~\ref{fig:vir_comparison} we compare the values of $\alpha_{\mathrm{vir}}$ obtained with the above methods and the method used in Section~\ref{sec:vir_parameter}. We find that $\alpha_{\mathrm{vir}}$ decreases with time for the pillars in all three methods, which suggests that our basic conclusions regarding the time evolution of the pillars is independent of the choice of method. However, we note that the values of $\alpha_{\mathrm{vir}}$ vary significantly with the method adopted, especially in the isolated, spherical-cloud method. This is because the definition of $\alpha_{\mathrm{vir}}$ in this method is based on global parameters, assuming a spherical, homogeneous cloud. This is far from realistic, as the pillars are highly non-homogeneous and non-spherical. The gas in the pillars is turbulent, and hence can be concentrated locally in fractal structures that cause a decrease in $\alpha_{\mathrm{vir}}$ without necessarily increasing the total mass in the region. In addition, the spherical cloud is assumed to be isolated, and hence the dynamical effects on the gravitational potential due to the gas outside the boundary of the pillar is not considered \citep{federrath_klessen_2012}. The large difference in the value of $\alpha_{\mathrm{vir}}$ obtained with this approximation suggests that it should be used and interpreted with caution, and may not necessarily represent the true dynamical state of the gas in the cloud.

\bsp	
\label{lastpage}
\end{document}